\documentclass[prl,twocolumn,floatfix,superscriptaddress]{revtex4-1}
\usepackage {graphicx} 
\usepackage {amssymb} 
\usepackage {amsmath}

\newcommand{\reffig}[1]{Fig.~\ref{#1}}
\newcommand{\refeq}[1]{Eq.~(\ref{#1})}

\DeclareMathOperator{\real}{Re}
\DeclareMathOperator{\imag}{Im}

\DeclareMathOperator{\erf}{erf}

\begin{document}

\title{ Tailoring THz radiation by controlling tunnel photoionization events in gases}

\author{I. Babushkin}
\email[Corresponding author: ]{babushkin@wias-berlin.de}
\affiliation{Weierstra\ss -Institut f\"ur Angewandte Analysis und Stochastik, 10117
Berlin, Germany}
\author{S. Skupin}
\affiliation{Max Planck Institute for the Physics of Complex Systems, 01187 Dresden,
Germany}
\affiliation{Friedrich Schiller University, Institute of Condensed Matter Theory and
Optics, 07743 Jena, Germany}
\author{A. Husakou}
\affiliation{Max-Born-Institut f\"ur Nichtlineare Optik und Kurzzeitspektroskopie, 12489
Berlin, Germany}
\author{C. K\"ohler}
\affiliation{Max Planck Institute for the Physics of Complex Systems, 01187 Dresden,
Germany}
\author{E. Cabrera-Granado}
\affiliation{Max Planck Institute for the Physics of Complex Systems, 01187 Dresden,
Germany}
\author{L. Berg\'e}
\affiliation{CEA-DAM, DIF, F-91297 Arpajon, France}
\author{J. Herrmann}
\affiliation{Max-Born-Institut f\"ur Nichtlineare Optik und Kurzzeitspektroskopie, 12489
Berlin, Germany}

\date{\today}
 
\begin{abstract}
Applications ranging from nonlinear terahertz spectroscopy to remote
sensing require broadband and intense THz radiation which can be
generated by focusing two-color laser pulses into a gas. In this
setup, THz radiation originates from the buildup of the
electron density in sharp steps of attosecond duration due to tunnel
ionization, and subsequent acceleration of free electrons in the laser
field. We show that the spectral shape of the THz pulses generated by
this mechanism is determined by superposition of contributions
from individual ionization events. This provides a straightforward 
analogy with linear diffraction theory, where the
ionization events play the role of slits in a grating. This analogy
offers simple explanations for recent experimental observations and
opens new avenues for THz pulse shaping based on temporal control
of the ionization events. We illustrate this novel technique by
tailoring the spectral width and position of the
resulting radiation using multi-color pump pulses.
\end{abstract}

\maketitle

\section{introduction  }
\label{sec:intro}

Recently, THz technology has attracted significant attention, because
it can provide unique analytical and imaging tools in nonlinear and
time-domain spectroscopy, remote sensing, biology and medicine,
security screening, information and communication technology (see
e.g.\cite{tonouchi07,sakai05,mittleman02,liu10b}). Among the various
THz sources, employing two-color ionizing femtosecond pulses in gases
\cite{kim08b,reimann08,cook00, kress04, bartel05, kresz06, xie06,
  thomson07, kim07,reimann07} stands out through absent damage
threshold, low phonon absorption and zero interface reflection. The
two-color scheme provides striking performance in terms of bandwidth
that can exceed 100 THz and thus even covers far- and mid-infrared, as
well as high peak fields up to the MV/cm range.

In the standard realization of this method a femtosecond laser pulse
and its second harmonic are collinearly focused into a gas, such as
ambient air, yielding powerful THz radiation with well-defined
electric field values.
Initially, this process has been explained by four-wave mixing
rectification via third-order nonlinearity
\cite{cook00,reimann07}. However, a threshold for THz generation
\cite{kress04,thomson07} as well as plasma current measurements
\cite{kim08b} indicated that plasma formation plays an important role
in the generation process. Besides, the Kerr nonlinearity is far too
small \cite{bartel05,babushkin10a} to explain the observed THz field
strength in such a setup.  In order to solve this discrepancy, Kim et
al. \cite{kim07,kim08b} have put forward the so-called photocurrent
mechanism. In this mechanism tunneling ionization and subsequent
electron motion produce a quasi-DC photoinduced current, which in turn
emits THz radiation. The quasi-DC plasma current can be efficiently
produced in a gas irradiated by two-color laser pulses
\cite{kim08b,reimann08,cook00, kress04, bartel05, kresz06, xie06,
  thomson07, kim07,reimann07}, in multicolor pump schemes
\cite{petersen10} and in chirped \cite{wang08b} or few-cycle pulses
\cite{wu08a,silaev09}.

Results published so far
\cite{thomson07,kim08b,kim09,petersen10,thomson10,babushkin10,babushkin10a,das10,rodriguez10}
provide several important features of the THz spectrum.  However, a
general framework for analysis, control and design of THz radiation is
still missing.  Similar to pulse shaping techniques in the optical
range, the broadband control of THz waveforms and its spectral
properties is highly desirable for many applications, while the
existing techniques are applicable only for a narrow bandwidth of few
THz \cite{ahn03,damico09,das10}.

In this paper we advance the understanding of the fundamental physics
in THz generation in gases as well as examine its basic mechanism
associated with plasma formation. We demonstrate that THz generation
in gases is intrinsically connected to optically-induced stepwise
increase of the plasma density due to tunneling ionization.  New
spectral components are emitted in a discrete set of attosecond-scale,
ultra-broadband bursts associated with these ionization events; in the
spectral representation THz emission results from an interference of
these radiation bursts.

Using appropriate field shapes of the pump pulses, e.g. tuning the
frequencies and phases of two- or multi-color pulses, the temporal
positions of the ionization events can be controlled. In turn, this
allows to design the emitted radiation within a broad spectral range
of more than 100 THz.
This new concept of prescribing the phase and amplitude of each
elementary contribution responsible for THz emission has a remarkable
analogy in linear diffraction theory: The produced spectrum can be
understood like the far-field diffraction pattern of a grating, where
the ionization events play the role of the slits.
A comprehensive (3+1)-dimensional non-envelope model of propagation is
used to verify by means of numerical simulations that the simple
picture drawn above captures the essential physics of THz generation.

  \section{Structure of the THz spectrum }
  \label{sec:struct}

\begin{figure*}
\includegraphics[width=\textwidth]{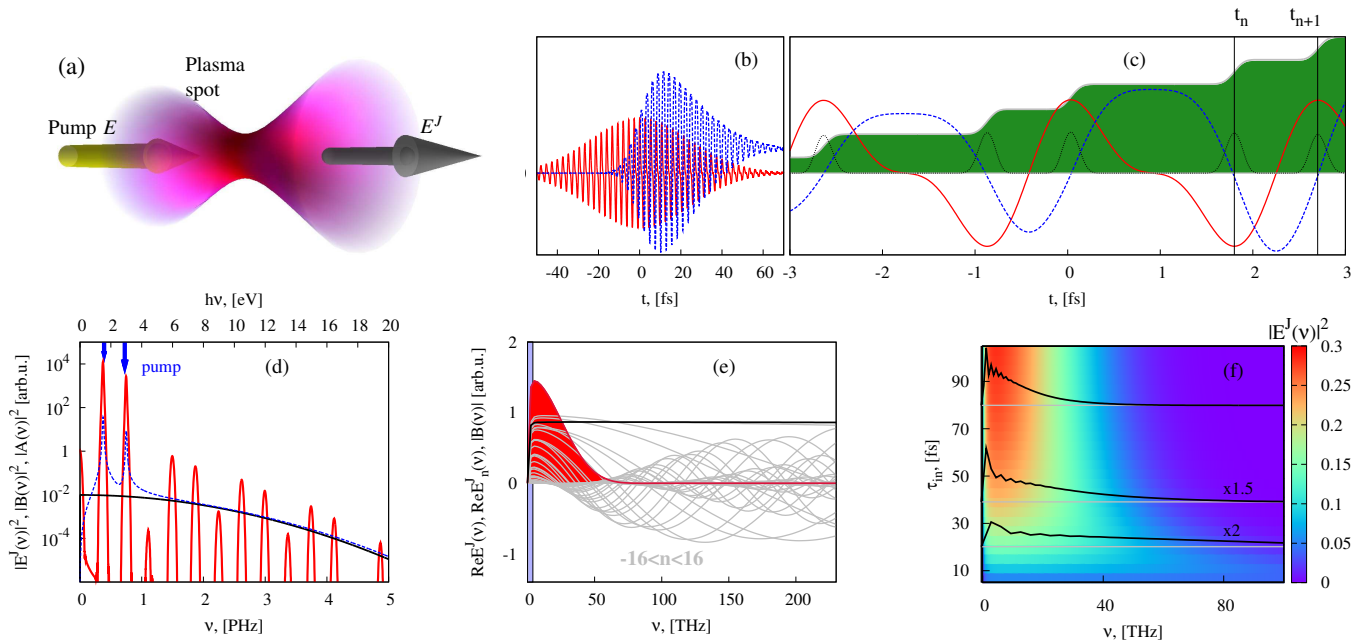}
\caption{ \label{fig:base} Mechanism and spectral properties of
  terahertz generation by ionizing two-color pulses. (a) Schematic
  setup, (b) two-color pump field $E(t)$ (red curve) and current
  $J(t)$ (blue curve) induced by tunneling-ionization, (c) step-wise
  modulation of the free electron density (filled green curve) and
  ionization rate $W(t)$ (thin dotted black curve), besides, few
  cycles of the electric field $E(t)$ (red curve) and the current
  $J(t)$ (blue curve) are shown.  (d) Spectrum $|E^J(\nu)|^2$ vs.\
  frequency $\nu=\omega/2\pi$ on a large frequency scale (red
  curve). Each ionization event at time $t_n$ induces a broadband
  radiation described by the form-factors $B(\nu)$ (black curve) and
  $A_n(\nu)$ (blue curve). The sum of all partial contributions in
  \refeq{eq:1} yields the resulting spectrum $E^J(\nu)$.  (e) Partial
  contributions $\real E_n^J(\nu)$ from the nth ionization event for
  $- 16<n<16$ (grey curves) and the total spectrum $\real E^J(\nu)$
  (filled red curve) in the THz to mid-infrared spectral region
  ($\imag E^J(\nu)$ is negligible here).  The black curve shows
  $B(\nu)$. The decay of $E^J(\nu)$ for $\nu \gtrsim 20$ THz is solely
  due to destructive interference of different $E^J_n(\nu)$. In
  contrast, in the shaded region ($\nu<5$~THz) $E^J(\nu)$ decays
  because of decrease of $B(\nu)$. (f) Spectrum $|E^J(\nu)|^2$ vs.\
  input pulse duration $\tau_{\rm in}$.  As the number of ionization
  events decreases with $\tau_{\rm in}$, the spectral width increases,
  as also confirmed by 3D simulations for $\tau_{\rm in}=20$, 40, and
  80~fs (black curves).
} \end{figure*}

  We start our analysis in the so-called local current (LC) limit,
  that is, we consider a small volume of gas irradiated by the strong
  pump field $E(t)$ [see \reffig{fig:base}(a)]. Electrons produced due
  to ionization build up a current $J(t)$ which, in turn, produces an
  electromagnetic wave $E^J(t) \propto \partial_t J(t)$ \cite{Jefimenko}. We focus here
  on the pump intensities where tunnel ionization is the dominant
  ionization process. In this regime, the free electron density $\rho(t)$
  increases stepwise in short attosecond-scale ionization events
  corresponding to maximum amplitudes of the pump field at times $t_n$
  [see \reffig{fig:base}(c)].  This stepwise increase of the free
  electron density $\rho(t)$ was recently confirmed experimentally
  \cite{uiberacker07,verhoef10}.  The ionization events are well
  separated from each other and thus $\rho(t)$ can be written as a sum
  over contributions from all ionization events: $\rho(t) = \sum_n
  \delta \rho_n H_n(t)$.  Each ionization event has a well defined
  amplitude $\delta \rho_n$ and shape $H_n(t)$.  Since all ionization
  events have almost the same shape [see green filled curve in
  \reffig{fig:base}(c)], one can assume $H_n(t) \simeq H(t-t_n)$,
  where $H$ is a ``smoothed'' step function.  An approximate
  analytical formula for $H(t)$ is given in the Methods.

In \reffig{fig:base}(b-e) an exemplary two-color pump field $E(t) =
\mathcal{E}(t)[\cos{(\omega_0 t)} + r \cos{(2\omega_0 t + \theta)}]$
is considered, where $\omega_0=2\pi\times 375$ ps$^{-1}$ is the
fundamental frequency (corresponding to the wavelength of 800 nm),
$r=\sqrt{0.2}$ and $\theta=\pi/2$ are relative phase and amplitude of
the second harmonic (SH) component. For such a two-color pulse (and
assuming $r<1$) the positions of ionization events are given by
$\omega_0 t_n \approx \pi n-2(-1)^nr\sin\theta$
\cite{babushkin10a}. $\mathcal{E}(t)$ is a Gaussian pulse envelope
shape with full-width at half-maximum (FWHM) pulse duration
$\tau_{in}=40$ fs.

With the assumption that electrons are born with zero initial
velocity, the plasma current $J(t)$ [blue curves in
Fig.~\ref{fig:base}(b,c)] is governed by
  \begin{equation}
    \label{eq:J}
\partial_tJ(t)+\gamma J(t)=\frac{q^{2}}{m}E(t)\rho(t),
\end{equation}
where $\gamma$ is the electron-ion collision rate ($\gamma \cong 5$
ps$^{-1}$ at atmospheric pressure) and $m,q$ are electron mass and
charge~\cite{kim07,babushkin10a}.  Then, in Fourier domain the
generated electromagnetic wave in the LC limit $E^J(\omega) \propto
\omega J(\omega)$ reads (see Methods)
\begin{equation}
  \label{eq:1}
  E^J(\omega) = \sum_n \left[A_n(\omega)- C_nB(\omega)\right]e^{i\omega t_n} .
\end{equation}
Here, $A_n(\omega) = qg \omega \delta \rho _{n}
\mathcal{F}[H(t)v_f(t+t_n)]$, $C_n = q\delta \rho _{n}v_{f}(t_{n})$,
and $B(\omega) = g \omega \mathcal{F}[H(t)e^{-\gamma t}]$, with
$\mathcal{F}$ denoting the Fourier transform, $v_f(t) = \frac{q}{m}
\int_{-\infty}^{t} E(\tau)e^{\gamma (\tau-t)}d\tau$ is the free
electron velocity, and $g$ is a constant.

\refeq{eq:1} allows us to identify the impact of each ionization
event.  The interpretation of \refeq{eq:1} is straightforward when we
consider the spectral dependence of the amplitudes $A_n(\omega)$ and
$B(\omega)$, which can be calculated analytically in a reasonable
approximation (see Methods):
\begin{equation}
  \label{eq:h}
  B(\omega)  \propto \frac{\omega}{\omega -
    i\gamma} \exp{\left( -\frac{\omega^2}{\sigma^2} \right)} ,
\end{equation}
where $\sigma \cong 10$ fs$^{-1}$ is the spectral width of a typical
ionization peak [see thin black curves in \reffig{fig:base}(c)].  In
contrast to $B(\omega)$, the amplitudes $A_n(\omega)$ depend
nontrivially on the pump pulse.
Examples of $A_n(\omega)$ and $B(\omega)$ are shown in
\reffig{fig:base}(d,e).  $B(\omega)$ features two clearly separated
frequency scales determined by $\gamma$ and $\sigma$. On the short
frequency scale $\omega \lesssim \gamma$ (up to 5~meV or $\sim
0-2$~THz for atmospheric pressure), $B(\omega)\propto\omega$, while
for larger frequencies 
$B(\omega)$ depends only
slowly on $\omega$ (on the scale of $\sim 10$ eV). 

If we consider the range from a few to approximately 100~THz,
\refeq{eq:1} can be significantly simplified. In this range
$B(\omega)$ can be considered as a constant, and additionally the
$A_n(\omega)$ are negligible as seen from Fig.~1(d).
\begin{equation}
  \label{eq:4}
  E^J(\omega) = B(\omega)\sum_n C_n e^{i\omega t_n}  \propto \sum_n C_n e^{i\omega t_n}.
\end{equation}

In other words, the resulting THz spectrum is a simple linear
superposition of the contributions $E^J_n=C_ne^{i\omega t_n}$ and its
structure is mainly governed by interference due to the spectral
phases $e^{i\omega t_n}$.  This interference is illustrated in
\reffig{fig:base}(e) [cf.\ also \reffig{fig:scont}(b)].
Note that in temporal domain \refeq{eq:4} corresponds to a set of
sharp peaks localized near $t_n$,  
each of them having the amplitude $C_n$ and the shape $\partial_t
\left[H(t-t_n)e^{-\gamma(t-t_n)}\right]$.  Thus, \refeq{eq:4}
describes sharp attosecond-long bursts of radiation localized near the
ionization events.

Further on, the same simple superposition principle is also valid for
higher frequencies in the more general \refeq{eq:1}. Indeed,
\reffig{fig:base}(d) shows that constructive spectral interference in
\refeq{eq:1} appears also for frequencies $\omega \sim l\omega_0$,
where $l$ is integer.  This finding is consistent with recent
experimental results~\cite{verhoef10}, where optical harmonics
generated due to tunneling-ionization-induced modulations of the
electron density were observed.  Therefore, our superposition
principle may also offer a complementary understanding of Brunel
radiation~\cite{brunel90,verhoef10}.

In fact, \refeq{eq:4} shows that THz generation can be understood
completely analogous to far-field interference patterns produced by a
diffraction grating \cite{born97}, where the ionization events play
the role of the slits.  Each of such ``slits'' produces a broad
``secondary wave'', and their interference results in narrow lines,
according to the Huygens-Fresnel principle \cite{born97}.  THz
radiation then corresponds to the zeroth diffraction order.  In
contrast to diffractive gratings, interference here takes place in the
frequency domain and not in position space.

In a somewhat oversimplified picture, one can explain the spectrum
shown in \reffig{fig:base}(d) analogous to interference patterns
produced by diffraction gratings of equidistant slits, $E^J(\omega)
\sim \sin(N \omega \delta t /2)/ \sin(\omega \delta t)$, where $N$ is
the number of ionization events and $\delta t \sim \pi/\omega_0$ is
the time interval between two subsequent ionization events (see
Auxiliary Material).
The width of the spectral line is inversely proportional to $N$, that
is, to the input pulse duration [see \reffig{fig:base}(f)]. Such
inverse dependence of the THz spectral width on the pump duration was
already observed experimentally \cite{bartel05}. It is also compatible
[black lines in \reffig{fig:base}(f)] with comprehensive 3D
simulations, where propagation and transverse effects are accounted
for (see Methods).

  \section{Control of the THz spectrum}
  \label{sec:control}

  Understanding the spectrum of the radiation emitted by the plasma
  current $J$ as an interference pattern provides a deeper insight
  into the mechanism responsible for the THz generation.  Beyond that,
  this understanding provides us with the possibility to shape the
  emitted radiation by suitable engineering of the pump field.  Here,
  we will restrict ourself to the THz and far-infrared domain
  described by \refeq{eq:4}. The latter indicates that, in order to
  control the THz spectral shape, we have to target the times of the
  ionization events $t_n$ and the values of $C_n=q\delta \rho_n
  v_f(t_n)$.

\begin{figure*}
\includegraphics[width=\textwidth]{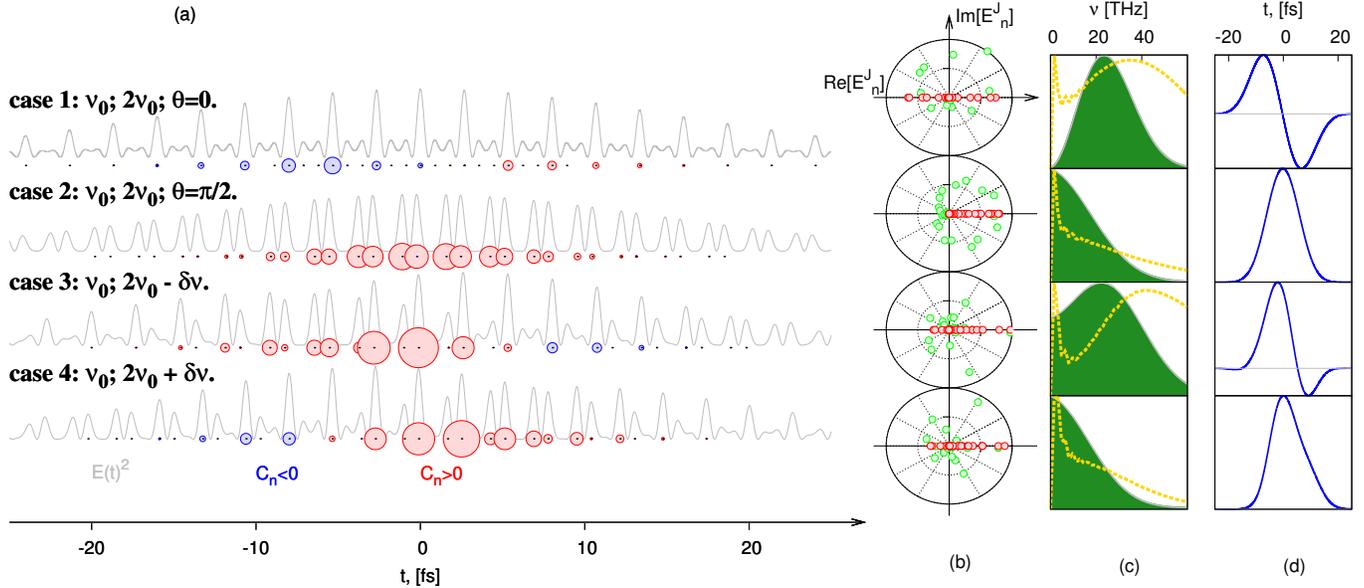}
\caption{ \label{fig:scont} Interference of contributions from
  different ionization events (a,b), THz spectrum (c) and temporal
  shape (d) for different values of the phase $\theta$ between the
  fundamental and second harmonics (cases 1, 2) as well as of the
  frequency shift $\delta \nu$ of the second harmonic (cases 3, 4;
  $\delta\nu=\mp 20$ THz, $\theta=\pi/4$).
(a)  Times of ionization events $t_n$ (center of circles) and amplitudes
  $C_n$ [radii of circles, red (blue) color indicates positive
  (negative) sign].  The gray curves show the electric field amplitude
  $E^2(t)$ of the pump pulse.  (b) Values of the summands in
  \refeq{eq:4} $E^J_n(\nu)$ in the complex plane for $\nu\cong0$ (red
  circles) and $\nu=25$~THz (green circles). For the cases (1) and (3)
  destructive interference occurs at $\nu \cong 0$ and is partially
  compensated at larger frequencies. (c),(d) Corresponding THz
  spectral shapes and temporal profiles obtained from \refeq{eq:4}
  ($\sum_n E_n^J$, filled green curve) and from 3D simulations at the
  linear focus $z=5$~mm (dashed yellow curves).  Note that the
  amplitudes for case (1) are about one order of magnitude smaller in
  comparison to the other cases.
} \end{figure*}

  Figure~\ref{fig:scont} shows several examples of THz generation from
  simple variations of the two-color scheme. Variations of the
  relative phase $\theta$ between fundamental and second harmonic
  component of the pump pulse have a strong influence on the THz
  spectral shape [see cases (1) and (2)], in addition to the known
  dependency of the THz yield~\cite{kim08b}.  This observation can be
  readily explained within our linear interference framework. For
  $\theta \cong 0$ (case 1) the coefficients $C_n$ are not sign
  definite [red and blue circles in \reffig{fig:scont}(a)]. Thus, for
  $\omega \cong 0$ different contributions in \refeq{eq:4} cancel each
  other, that is, destructive interference dominates.  Nevertheless,
  with increasing frequency $\omega$ the destructive interference is
  (partially) compensated by the spectral phases $e^{i\omega t_n}$.
  This behavior is shown in \reffig{fig:scont}(b), where contributions
  from the ionization events $E^J_n$ are visualized in the complex
  plane for different frequencies $\omega$. Consequently, the
  spectral maximum is shifted from zero [see \reffig{fig:scont}(c)].
  In contrast, for $\theta=\pi/2$ (case 2) all coefficients $C_n$ are
  positive.  This obviously ensures that at $\omega \cong 0$
  constructive interference dominates, and the phase factor
  $e^{i\omega t_n}$ can lead only to destructive interference when
  $\omega$ is increased. Therefore, the spectral maximum is located at
  zero frequency in this case. Moreover, the coefficients $C_n$ are
  much larger in amplitude, which explains the increased THz yield for
  $\theta=\pi/2$.


We verified this behavior by comprehensive (3+1)-dimensional (3D)
numerical modeling (see Methods), which was recently successfully
compared to experiments~\cite{babushkin10a}.  We consider a Gaussian
input beam with width $w_0=100~\mu$m and duration (FWHM) $\tau_{\rm
  in}=40$~fs for the 800~nm fundamental pump pulse (pulse energy $\sim
300~\mu$J).  The energy of the second harmonic at 400~nm is chosen as
7.5~\% of the fundamental, with duration and width smaller than the
values for the fundamental by a factor $\sqrt{2}$ (similar to typical
experimental conditions). These pulsed input beams are focused with
$f=5$~mm into argon at atmospheric pressure.  In contrast to the
simple LC model, the phase angle $\theta$ between the two components
shifts during propagation. For our particular configuration, an
initial value $\theta_{\rm in}\cong\pi/4$ guarantees maximum THz
energy yield of $\sim1.8~\mu$J; $\theta_{\rm in}\cong-\pi/4$ leads to
a minimum yield of $0.25~\mu$J; the generic value $\theta_{\rm in}=0$
yields $\sim1~\mu$J. THz spectra obtained for the two extremal cases
are shown in \reffig{fig:scont}(c), cases (2) and (1),
respectively. They agree remarkably well with the predictions obtained
from \refeq{eq:4}.

We also observe a high sensitivity of the THz spectra towards detuning
of the relative frequency ratio between the two pump components [see
case (3) and (4) in \reffig{fig:scont}]. Such pulses with so-called
incommensurate SH (frequency $2\nu_0$ is shifted by $\delta \nu$)
offer a straightforward control of the position of the maximum THz
spectral intensity, without decreasing the total THz yield. Recently,
this shift of the THz spectrum with $\delta \nu$ was predicted
\cite{kim09} and demonstrated experimentally~\cite{thomson10}.
Again, we can use our linear interference framework to explain this
effect. The responsible shifts of the ionization events can be
computed approximatively (see Auxiliary material) as $\omega_0
t_n\approx n\pi-2r(-1)^{n}\sin{( \theta \pm \pi n \delta \nu /
  \nu_0)}$.  The SH frequency shift $\pm \delta \nu$ makes the
effective phase angle $n$ dependent, and thus shifts the ionization
events in time. Using our diffraction-grating analogy, the effect
corresponds to increasing or decreasing the slit-distance along the
grating.  This frequency shift can be also calculated analytically
under some further approximations (see the Auxiliary Material).

\begin{figure}
\includegraphics[width=0.5\textwidth]{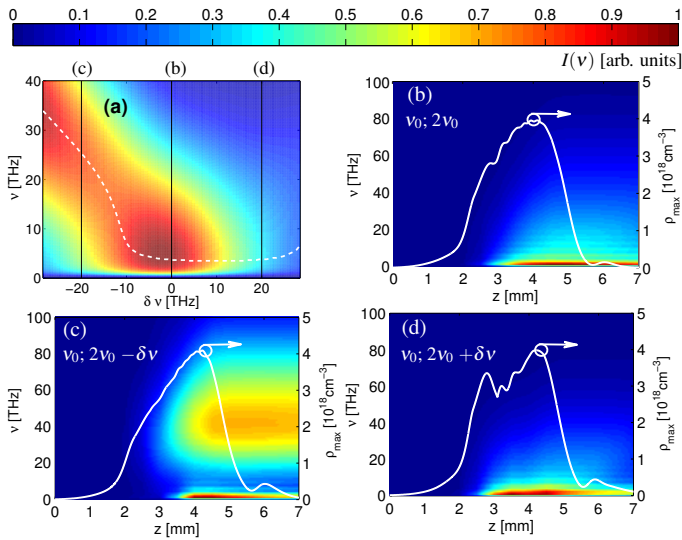}
\caption{ \label{fig:two-col}
  THz generation by two-color pump pulses with incommensurate
  frequencies.  (a) THz spectrum vs. frequency shift $\delta \nu$ of
  the second harmonic computed from \refeq{eq:1}. The dashed white
  line shows the position of the spectral maximum.  (b),(c),(d) THz
  spectrum (color plot) as well as the maximum plasma density (white
  line) along the propagation axis according to 3D simulations for
  $\delta \nu = 0$ and $\pm20$~THz, respectively.  } \end{figure}

Because the coefficients $C_n$ are not sign definite [see
\reffig{fig:scont}(a,b), the ``secondary waves'' are not in-phase],
such non-equidistant grating can lead to a spatial shift of the zeroth
diffraction order, i.e., a frequency shift of the maximum THz spectral
intensity.  As it is seen from \reffig{fig:scont}(c) and
\reffig{fig:two-col}(a) the THz frequency shift is not symmetric with
respect to the sign of $\delta \nu$.  This effect occurs for pulses
with finite duration (see Auxiliary Material).

Again, our predictions from \refeq{eq:4} are in good agreement with 3D
numerical simulations.  Incommensurate two-color pulses clearly
produce the predicted THz spectral shift for negative SH detuning [see
\reffig{fig:two-col}(c)].  Moreover, neither linear nor nonlinear
propagation effects of the pump pulse due to focusing or laser-induced
plasma seem to jeopardize this effect. The energy in the shifted THz
pulse is $\sim1~\mu$J, like in the corresponding commensurate case
[see \reffig{fig:two-col}(b)]. On the other hand,
\reffig{fig:two-col}(d) reveals that, as expected, a positive SH
detuning does not alter the THz spectral shape significantly, but the
energy yield reduces by a factor of two to $\sim0.5~\mu$J.

\begin{figure}
\includegraphics[width=0.5\textwidth]{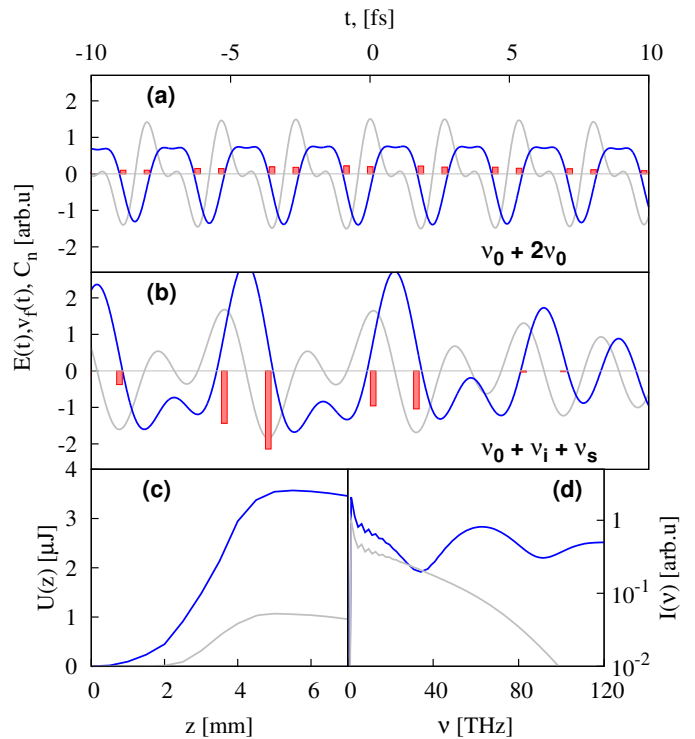}
\caption{ \label{fig:opa}
  Comparison of the THz characteristics produced by two-and
  three-color pump pulses.  (a) Field shape $E(t)$ (gray line), free
  electron velocity $v_f(t)$ (blue line) as well as ionization
  amplitudes $C_n$ (bars at ionization events) for the two-color pump
  (see \reffig{fig:base}).  (b) Same quantities for the three-color
  pump, e.~g., for an OPA with frequencies $\nu_0=\nu_{\rm s}+\nu_{\rm
    i}$  ($\nu_{\rm s}=0.55\nu_0$, $\nu_{\rm i}=0.45\nu_0$). (c) THz
  energies $U$ vs.\ propagation distance $z$ according to 3D
  simulations and (d) corresponding spatially integrated THz spectra
  at the position of the linear focus $z=5$~mm for three-color (blue
  lines) and two-color (gray lines) cases. }
\end{figure}

Let us finally consider a more complex example of THz spectral
engineering. Namely, we attempt to significantly increase the spectral
width by using a field shape which optimizes the ``amplitudes'' $C_n$
of just a few ionization events, suppressing the others. This leads to
THz-to-mid-infrared supercontinuum, which is of great importance for
many applications. For this purpose, we resort to a three-color pump
setup involving an optical parametric amplifier (OPA).  We assume a
fundamental frequency $\nu_0=375$~THz ($\lambda_0 = 800$~nm) and
signal and idler frequencies $\nu_{\rm s}$ and $\nu_{\rm i}$ obeying
$\nu_{\rm s} + \nu_{\rm i} = \nu_0$  ($\nu_{\rm s}=0.55\nu_0$,
$\nu_{\rm i}=0.45\nu_0$). 
 From \reffig{fig:opa}(b), one
can see that in such a configuration only a few ionization events have
large amplitudes $|C_n|$.  This immediately leads to a much broader
THz spectrum, as it is obvious from the above discussed diffraction
grating analogy.
In \reffig{fig:opa}(c,d) the results of 3D simulations for this case
are shown, assuming that each of the three pump components contains one
third of the total pulse energy of 300 $\mu$J while all other
parameters are kept unchanged.

Quite interestingly, the three-color configuration we use provides
much larger values for the $|C_n|$ than the previous two-color scheme
with comparable pump energy [cf.\ \reffig{fig:opa}(a) and (b)].  In
the LC limit, the resulting THz yield is about 40 times larger. In our
simulations, this increased THz yield is indeed visible in the
beginning of the propagation $z<2.5$~mm.  However, upon further
propagation saturation effects, mainly coming from plasma defocusing
of the pump and depletion of neutral atoms, limit the total THz energy
to a few $\mu$J.

 \section{Conclusion and Outlook }
  \label{sec:concl}

  In summary, we have shown that THz emission in gases by multi-color
  laser fields is intrinsically connected with the attosecond jumps of
  the plasma density induced by tunneling ionization. The THz spectrum
  is described by a superposition of the spectrally ultrabroad
  contributions from separate ionization events.

  This approach explains basic features of THz radiation such as the
  inverse dependence of the spectral width on the pulse duration and
  the shift of THz spectral maxima for non-integer frequency offsets
  of two-color pumps.
  Moreover it allows to manipulate the THz waveform and to shape the
  THz spectrum by controlling the tunneling photo-ionization. In the
  present article we demonstrated only basic features of such a
  control, using frequencies and phases of two- or three-colour pump
  fields. We note that other degrees of freedom of light such as
  polarization or spatial shape of the pump pulses can also be used
  for THz control. Noteworthy, our approach is not limited to THz,
  far- and mid-infrared range but can be potentially extended to
  frequencies much higher than the pump one, i.e., to optical
  harmonics.

  We believe that this novel strategy to control fundamental
  properties of THz pulses could become a basic tool in optical
  technology: a ultra-broadband pulse shaper working in the range from
  THz to mid-infrared. Just as optical pulse shapers for femtosecond
  pulses are indispensable in ultrafast physics, the proposed THz
  control can find promising applications in many fields of research,
  including, among others, THz time-domain spectroscopy, chemical
  reaction dynamics, and optical signal processing.


\appendix

\section{Methods}
\label{sec:methods}


\subsection{Discrete ionization model}
\label{sec:model-descr-ioniz}

The electron density $\rho(t)$ is described by the equation 
\begin{equation}
\label{eq:rho}
\partial_t \rho=W_{\mathrm{ST}}(E)[\rho _{at}-\rho(t)],
\end{equation}
where $\rho _{at}$ denotes the neutral atomic density.  We use a
quasi-static tunneling ionization rate for hydrogen-like atoms
\cite{thomson07,babushkin10a}: $W_{\mathrm{ST}}(E)=4\omega
_{a}(r_{H})^{2.5}[E_{a}/|E|]\exp [-2(r_{H})^{1.5}E_{a}/3|E|]$, where
{$E_{a} = m^{2}q^{5}/(4 \pi \epsilon_0)^3\hbar ^{4}$, $\omega
  _{a}=mq^{4}/(4 \pi \epsilon_0)^2 \hbar ^{3}$, and
  $r_{H}=U_{Ar}/U_{h}$}. $U_{h}$ and $U_{Ar}$ are the ionization
potentials of hydrogen and argon, respectively.

Supposing that the ionization process occurs only near maxima of the
pump electric field at times $t_n$, one finds $E(t) \approx E_0 +
E_2(t-t_n)^2$ with $E_2=\partial_{tt} E(t_n)/2$ and $E_0 =
E(t_n)$. Then, assuming $\rho _{at}\gg \rho$, we can evaluate
$W_{\mathrm{ST}}(E) \approx W_{\mathrm{ST}}[E(t_n)]
\exp\left\{-\sigma^2(t-t_n)^2/4\right\}$ where $\sigma^2=
8(r_{H})^{1.5}E_{a}|E_2|/3|E_0|^2$, and find
\begin{equation}
\rho(t) = \sum_n \delta
\rho_n H(t-t_n), \label{eq:2}
\end{equation}
with
$$
H(t-t_n) \propto
\int\limits_{-\infty}^{t}e^{-\frac{\sigma^2(\tau-t_n)^2}{4}}d\tau =
\sqrt{\frac{\pi}{\sigma^2}}\left\{1+\erf\left[\frac{\sigma(t-t_n)}{2}\right]\right\}.
$$
Strictly speaking, $\sigma$ is different for every ionization
event. The difference, however, is typically less than a few percent,
so we assume here a ``typical'' value $\sigma \cong 10$ fs$^{-1}$.

The free electron current $J$ is given by 
$$
J(t)=q\int_{-\infty}^{t}\left[\partial_\tau{\rho}(\tau)\right]\left[v_f(t)-e^{\gamma
    (\tau-t)}v_f(\tau)\right]d\tau,
$$ 
which is equivalent to \refeq{eq:J} (see
\cite{kim07,babushkin10,babushkin10a}).  Using \refeq{eq:2}, the LC
$J$ can be written as a sum over all contributions from separate
ionization events $J(t)=\sum_n J_n(t)$, and hence
\begin{equation}
J(t) = \sum_n q\delta \rho _{n} H(t-t_n)\left\{v_{f}(t)-e^{\gamma
    (t_n-t)}v_{f}(t_{n})\right\}.
\label{eq:j1}
\end{equation}
In the LC limit, assuming that the plasma source is contained in a small
volume $\Delta V$, the field produced by the plasma current
$J(\omega)$ in frequency domain is $E^J=g\omega J(\omega)$ \cite{Jefimenko}, where $g =
i \frac{\omega \Delta V}{4 \pi \epsilon_0 c^2 r} \exp{(i \omega
  \frac{r}{c})}$, $c$ is the speed of light, $\varepsilon_0$ is the
vacuum permittivity and $r$ is the distance between the plasma source
and the point where the field is measured.

Thus, in order to obtain \refeq{eq:1}, we have to multiply the Fourier
transform of \refeq{eq:j1} by $\omega g$.  Further on, by performing
the Fourier transform of $H(t) \exp{(-\gamma t)}$ and neglecting a
slowly varying phase term $\exp{(-2i\gamma\omega/\sigma^2)}$ we obtain
\refeq{eq:h}.

\subsection{Comprehensive numerical model}
\label{sec:3d-model}

We use the unidirectional pulse propagation equation \cite{kolesik04,
  babushkin10a} for linearly polarized pulses
\begin{equation}
  \partial _{z}E(\omega)=i\sqrt{k^2(\omega )-k_{x}^{2}-k_{y}^{2}}E(\omega)+i%
  \frac{\mu _{0}\omega ^{2}}{2k(\omega )}\mathcal{P}_{\mathrm{NL}}(\omega).
\label{Eq.(1)}
\end{equation}%
Here, $E(k_{x},k_{y},z,\omega )$ is the relevant electric field
component in Fourier domain with respect to the transverse coordinates
$(x,y)$ and time, $k=\omega n(\omega )/c$ is the wavenumber, $c$ is
the speed of light and $n(\omega)$ is the linear refractive index of
argon~\cite{dalgarno60}.  The nonlinear polarization
$\mathcal{P}_{\mathrm{NL}}(\omega)=P_{\mathrm{Kerr}}(\omega)+iJ(\omega)/\omega
+iJ_{\mathrm{loss}}(\omega)/\omega $ accounts for third-order
nonlinear polarization $P_{\mathrm{Kerr}}(t)$, electron current $J(t)$
and a loss term $J_{\mathrm{loss}}(t)$ due to photon absorption during
ionization.
In our 3D numerical code, Eqs.~(\ref{Eq.(1)}), (\ref{eq:J}), and
(\ref{eq:rho}) are solved using a standard spectral operator splitting
scheme.  Particular attention is paid to resolve all relevant time
scales, from tens of attosecond (ionization steps) to a few picosecond
(THz radiation).
Although, strictly speaking, \refeq{Eq.(1)} does not describe the
propagation of waves below the plasma frequency properly, this plays
only a minor role because the THz radiation is emitted from the
leading plasma front and thus does not propagate in the plasma
\cite{koehler11}. Spectra obtained using \refeq{Eq.(1)} were found to
be in good agreement with experimental results \cite{babushkin10a}.


\section{Auxiliary Material}
\label{sec:auxiliary-material}

We consider for simplicity the case of a two-color pulse with duration $T_0$ having
the square-shaped envelope $E(t)=0$ for $|t|>T_0/2$ and
 \begin{equation}
   \label{eq:11}
 E(t) = A_1 \cos\{\omega_0 t\} + A_{2}
  \cos\{(2\omega_0 + \delta \omega) t + \theta \}
 \end{equation}
 for $|t|\le T_0/2$. Here  $\delta \omega$ is  a detuning from the second
 harmonic, which is assumed to be small compared to the main frequency
 ($\delta \omega \ll \omega_0$).  We also neglect the plasma decay $\gamma$ and assume $r \equiv A_{2}/A_{1} \ll 1$. 

To determine the positions  $t_n$ of the tunneling ionization events
(maxima of the electric field) along the time axis we equate the time
derivative of \refeq{eq:11} to zero,
set $t_n=t_{n0}+\delta t_n$ and expand the resulting expression up to first orders in $\delta\omega$ and $\delta t_n$, yielding
 \begin{gather}
 \notag
   \omega_{0}t_{n}\approx n\pi- 2r(-1)^{n}\left(n\pi \delta \omega/\omega_0
     \cos\theta + \sin\theta\right) \\ \approx n\pi- 2r(-1)^{n} \sin(\theta +
   n\pi \delta\omega/\omega_0),   \label{eq:13} 
 \end{gather}
where 
it is assumed that $\pi n\delta\omega/\omega_0 \ll 1$. Using
Eqs.~(\ref{eq:11}), (\ref{eq:13}) and neglecting again
higher-order terms, we find
\begin{equation}
  \label{eq:6}
    v_f(t_n) = \frac{q}{m} \int \limits_{-\infty}^{t_n} E(\tau)d\tau \approx a_1 + na_2,
\end{equation}
where
\begin{gather}
  \label{eq:13b}
  a_{1}=-3qA_{1}r\sin \theta /2m\omega _{0}, \\
  \label{eq:13c}
  a_{2}=-3q\pi A_{1}r\delta\omega \cos \theta /2m\omega_{0}^2.
\end{gather}
We further assume
that the amplitude $\delta \rho_n$ of each step in the modulation
of the plasma density does not depend on the index $n$. For the
square-shaped pulse envelope considered here this essentially means that
saturation of the free electron density is neglected.
Then, according to
Eq.~(4) of the main article
\begin{equation}
  \label{eq:12}
E^J(\omega) \propto  (a_1S_1+a_2S_2),
\end{equation}
where
\begin{gather}
  \label{eq:s1}
 S_1 = \sum\limits_{n} e^{i \omega
t_{n}},\\
  \label{eq:s2}
S_2 = \sum\limits_{n} ne^{i \omega
t_{n}}. 
\end{gather}
Here, the summation runs over all ionization events.  Following
\refeq{eq:13}, the exponent $e^{i\omega t_n}$ can be expressed as
$e^{-2ir \tilde \omega \sin\theta+ing_+\tilde \omega}$ for even $n$
and $e^{2ir \tilde \omega \sin\theta-ing_-\tilde \omega}$ for odd $n$,
where $\tilde \omega = \omega/\omega_0$ is the normalized frequency
and $g_+ = \pi (2r\delta\omega\cos\theta/\omega_0+1)$, $g_- = \pi
(2r\delta\omega\cos\theta/\omega_0-1)$.  Considering an odd number of
ionization events $N$, 
being much larger than unity, 
we calculate the
expressions in \refeq{eq:12} by performing the summation from $-M$ to
$M$, where $M=(N-1)/2$. The term $S_1$ in \refeq{eq:s1} comprises
simple geometric series and gives after separate summation for odd and
even $n$:
\begin{equation}
    \label{eq:14}
S_1  \approx   e^{2ir\tilde
     \omega \sin\theta}\frac{\sin\left(M g_+ \tilde\omega
     \right)}{\sin\left(g_+ \tilde\omega\right)} + 
   e^{-2ir\tilde
     \omega \sin\theta} \frac{\sin\left(Mg_-\tilde\omega\right)}{\sin\left(g_-\tilde\omega\right)},
 \end{equation}
where $M\gg1$, so that $M+1\approx M$. 
A similar analysis is possible
for 
the term $S_2$ in \refeq{eq:s2}. 
As before, one performs the sum for odd and even values of $n$ separately.
We are then left with an arithmetico-geometric series 
\cite{1}, 
i.e., for arbitrary $q$
and integers $m_1,m_2$: 
    \begin{equation}
      \label{eq:4aux}
      \sum \limits_{n=m_1}^{m_2} nq^{n} = \frac{m_1(q-1)+q+q^{m_1+m_2+1}[m_2(q-1)-1]}{q^{m_1}(q-1)^2}.
    \end{equation}
After substituting Eq.(\ref{eq:4aux}) into Eq.~(\ref{eq:s2}) we find
      \begin{gather}
        \nonumber
        S_2 = \frac{i}{2}
        \left. 
        \biggl\{
     \frac{(M+2)\sin\left[Mg_-\tilde\omega\right]-
            M\sin\left[(M+2)g_-\tilde\omega\right]}{\sin^2\left(g_-\tilde\omega\right)}
          \right.  \\ \nonumber \times \exp{(-2ir\tilde \omega
            \sin\theta)}  + \\ + 
          \frac{(M+1)\sin\left[(M-1)g_+\tilde\omega\right]-(M-1)
            \sin\left[(M+1)g_+\tilde\omega\right]}{\sin^2\left(g_+\tilde\omega\right)} 
           \nonumber \\ \times
          \left. \exp{(2ir\tilde \omega \sin\theta)} \biggr\}. \right.  
        \label{eq:17}
      \end{gather}
      Important features can be refound from the above expressions.  In
      particular, for $\delta \omega =0$ we have $a_2=0$ and hence $E^J
      \propto S_1$.  In addition, $g_+=-g_-=\pi$ and thus, by introducing
      $\delta t = \pi/\omega_0$ we obtain:
      \begin{equation}
        \label{eq:1aux}
        S_1 \approx 2\cos{\left(2r\sin\theta
     \frac{\omega}{\omega_0}\right)}\frac{\sin\left( M \omega
      \delta t\right)}{\sin\left(\omega \delta t\right)}.
      \end{equation}
      The equation given at the end of Sec.~II in the
      main article is obtained from \refeq{eq:1aux} by assuming
      $\theta=0$ and $N \gg 1$, that is,
      $M=(N-1)/2\approx N/2$.

      For nonzero detuning $\delta \omega$ we can at least
      qualitatively reproduce the behavior presented in Fig.~2 (case
      3) and in Fig.~3(a) of the main article, i.e., the shift of the
      maxima of the THz spectra from zero frequency. The frequency
      shift obtained from \refeq{eq:12}, \refeq{eq:14}, \refeq{eq:17}
      is, however, symmetric with respect to the sign of $\delta
      \omega$, in contrast to the full LC model. The asymmetry
      reported in the main text [see, e.g., Fig. 1(d)] disappears with
      increasing of the pulse duration and thus is induced by the
      presence of the Gaussian pulse envelope.


\begin{figure}
\includegraphics[width=0.5\textwidth,clip=true]{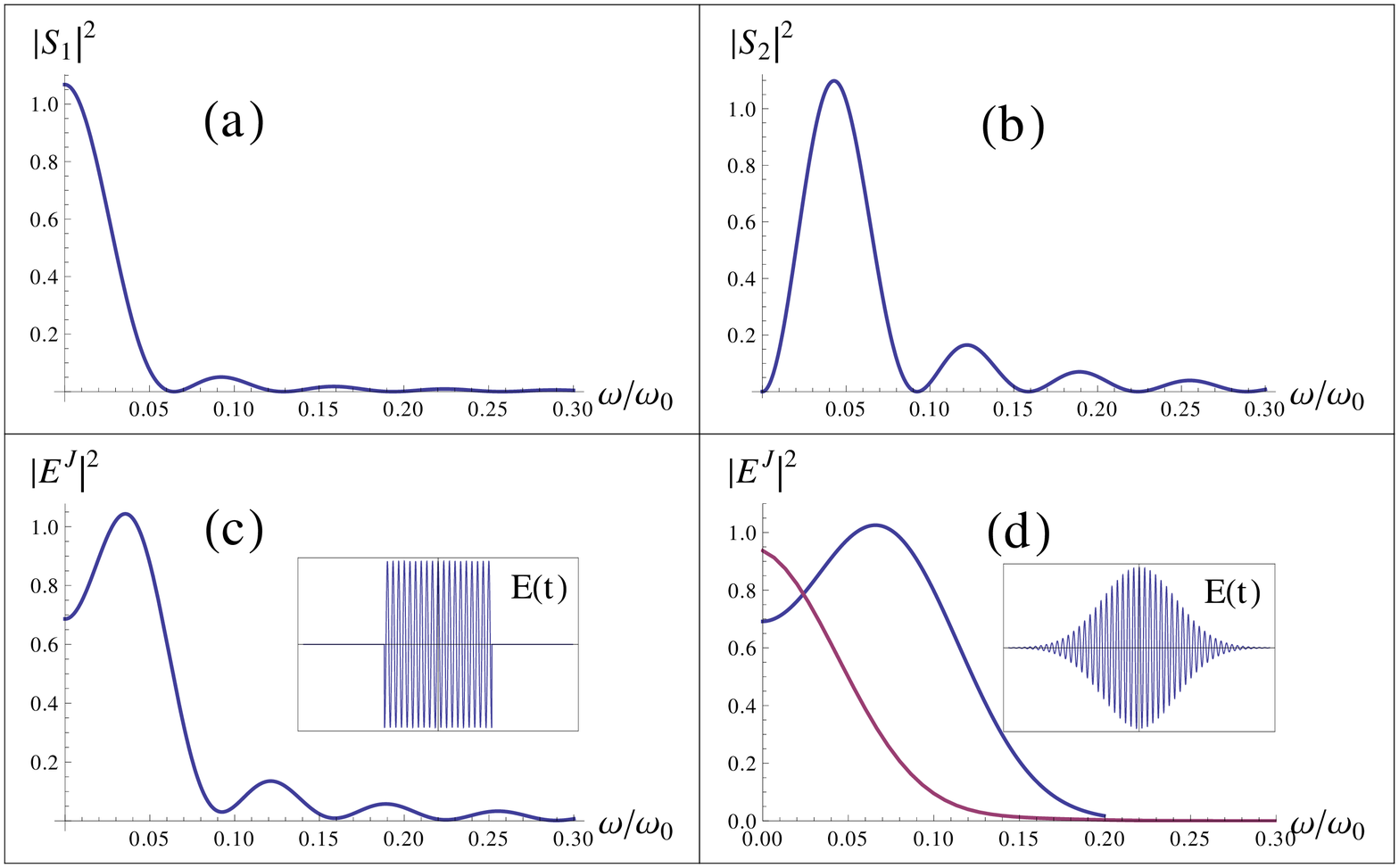}
\caption{ \label{fig:opaaux}
    Typical form factors (a) $|S_1|^2$ , (b) $|S_2|^2$ and (c) the
    field $|E^J|^2$ depending on $\tilde \omega = \omega/\omega_0$
    obtained from \refeq{eq:12}, \refeq{eq:14}, \refeq{eq:17} for
    $\theta=\pi/4$, $M=15$, $r=\sqrt{0.2}$, $\delta\omega/\omega_0=0.0533$
    ($\delta \nu=20$ THz). For comparison, in (d) the spectral shapes
    from Fig. 2(c) of the main article are repeated [blue curve
    corresponds to the case 3 ($\delta \nu =-20$ THz) and red curve to
    the case 4 ($\delta \nu=20$ THz)].  The insets in (c) and (d) show
    the corresponding pump pulse shapes.
}
\end{figure}




\end{document}